\documentclass[reprint,amsmath,amssymb,pra]{revtex4-1}
\usepackage{graphicx}
\usepackage{xcolor}
\usepackage{amsmath,amssymb}
\usepackage{upgreek}
\usepackage{verbatim}
\usepackage{comment}

\usepackage{hyperref}

\newcommand{\VL}{V_\text{L}}
\newcommand{\VR}{V_\text{R}}

\newcommand{\VI}{V_\text{I}}
\newcommand{\VQ}{V_\text{Q}}

\newcommand{\deltaQ}{\delta Q}
\newcommand{\DeltaQ}{\Delta Q}
\newcommand{\Qtilde}{\tilde{Q}}

\newcommand{\IB}{I_\text{B}}
\newcommand{\IFlux}{I_{\Phi}}

\begin{document}

\title[Reflectometry]{Sensitive radio-frequency read-out of quantum dots using an ultra-low-noise SQUID amplifier}

\author{F.~J.~Schupp$^{1,\dagger}$}

\author{F.~Vigneau$^{1,\dagger}$}

\author{Y.~Wen$^1$}

\author{A.~Mavalankar$^1$}

\author{J.~Griffiths$^2$}

\author{G.~A.~C.~Jones$^2$}

\author{I.~Farrer$^{2,3}$}

\author{D.~A.~Ritchie$^2$}

\author{C.~G.~Smith$^2$}

\author{L.~C.~Camenzind$^4$}

\author{L.~Yu$^4$}

\author{D.~M.~Zumb\"{u}hl$^4$}

\author{G.~A.~D.~Briggs$^1$}

\author{N.~Ares$^1$}

\author{E.~A.~Laird$^{1,5}$}
\email{e.a.laird@lancaster.ac.uk\\$^\dagger$These authors contributed equally to this work}

\affiliation{$^1$University of Oxford, Department of Materials, 16 Parks Road, Oxford OX1 3PH, UK\\
$^2$Cavendish Laboratory, J.J. Thomson Avenue, Cambridge CB3 0HE, UK\\
$^3$Department of Electronic and Electrical Engineering, University of Sheffield, Sheffield S1 3JD, UK\\
$^4$Department of Physics, University of Basel, 4056 Basel, Switzerland \\
$^5$Department of Physics, University of Lancaster, Lancaster, LA1 4YB, United Kingdom}

\keywords{radio frequency, reflectometry, SQUID amplifier, capacitance sensing}

\date{\today}
\begin{abstract}
Fault-tolerant spin-based quantum computers will require fast and accurate qubit readout. This can be achieved using radio-frequency reflectometry given sufficient sensitivity to the change in quantum capacitance associated with the qubit states. Here, we demonstrate a 23-fold improvement in capacitance sensitivity by supplementing a cryogenic semiconductor amplifier with a SQUID preamplifier. The SQUID amplifier operates at a frequency near 200~MHz and achieves a noise temperature below 600~mK when integrated into a reflectometry circuit, which is within a factor 120 of the quantum limit. It enables a record sensitivity to capacitance of $0.07\,\text{aF}/\sqrt{\text{Hz}}$. The setup is used to acquire charge stability diagrams of a gate-defined double quantum dot in a short time with a signal-to-noise ration of about 38 in $1\,\upmu\text{s}$ of integration time.
\end{abstract}
                             
\maketitle

\section{Introduction}
Electron spins in semiconductors are among the most advanced qubit implementations, and are a potential basis of scalable quantum computers fabricated using industrial processes~\cite{DiVincenzo2000,Veldhorst2016a,Vandersypen2017}. A useful computer must correct the errors that inevitably arise during a calculation, which requires high single-shot qubit readout fidelity~\cite{Fowler2012}. The full surface code for error detection requires approximately half the physical qubits to be read out in every clock cycle of the computer~\cite{Terhal2015}. Until recently, single-shot readout in spin qubit devices could only be achieved via spin-to-charge conversion, detected by a nearby single-electron transistor (SET) or quantum point contact (QPC) charge sensor~\cite{Hanson2007,Barthel2009,Gaudreau2011,Veldhorst2015}.
However, the hardware is simpler and smaller if it uses dispersive readout, which exploits the difference in electrical polarizability between the singlet and triplet spin states in a double quantum dot~\cite{Petta2005,Petersson2010,Gonzalez-Zalba2015, Crippa2019}. The resulting capacitance difference between the two qubit states can be monitored via a radio-frequency (RF) resonator bonded to one of the quantum dot electrodes. Similar dispersive shifts also occur at charge transitions in the quantum dots, such that the reflected signal assists with tuning to the desired electron occupation~\cite{Crippa2017, volk2019,Watson2019}.
Dispersive readout has the advantage that it does not require a separate charge sensor, but often the capacitance sensitivity is insufficient for single-shot qubit readout even in systems with a long spin decay time~\cite{Jung2012,Ares2016,Colless2013,Hile2015,Betz2015a,Ahmed2018,Higginbotham2014a}. Recently, there have been demonstrations of dispersive single-shot readout in double quantum dot based systems~\cite{Pakkiam2018,West2018, Meunier2019, Vandersypen2019, Nichol2019}, but higher sensitivities are still desirable for improved readout fidelity.

High sensitivity also makes it possibly to rapidly measure charge stability diagrams and therefore speeds up quantum dot tuning.  For example, fast measurement has enabled video-mode tuning using an RF setup attached to one of the electrodes of a double quantum dot~\cite{Stehlik2015}. Auto-tuning techniques~\cite{Lennon2018, Moon2020, Nina2020}, which are often limited by measurement time, also benefit from increased measurement speed.

High-fidelity readout, whether dispersive or using a charge sensor, relies on low-noise amplifiers to attain good capacitance sensitivity.
Radio-frequency experiments until now have used semiconductor amplifiers cooled to $\sim$4~K.
Even lower noise can be achieved using amplifiers based on superconducting quantum interference devices (SQUIDs).
At microwave frequencies, Josephson parametric amplifiers (JPAs) and travelling wave parametric amplifiers approach the quantum limit of sensitivity~\cite{Castellanos-Beltran2007,Yamamoto2008,Macklin2015}.
Such amplifiers allow rapid measurements of charge parity in a double quantum dot~\cite{Stehlik2015}.
However, the JPAs previously used for quantum dot readout have a linear amplification range limited to an input power of $-130 \,\text{dBm}$ ~\cite{Stehlik2015, Morton2019}, they require a circulator inside the cryostat and a dedicated pump oscillator, and they are not commercially available. Most JPAs are optimized for a microwave frequency range well above 1~GHz, although operation as low as $650~\text{MHz}$ has been demonstrated~\cite{Vesterinen2017, Morton2019}. However, a lower frequency is desirable because the charge dipole in a singlet-triplet qubit only responds adiabatically to changes in the electric field if the interdot tunnel rate is much larger than the read-out frequency. If, on the other hand, the readout frequency approaches the inter-dot tunnel rate, the quantum capacitance is suppressed and read-out times increase~\cite{Chorley2012}. Increasing the tunnel rate leads to inelastic spin relaxation~\cite{Johnson2005}, and it is therefore usually limited to no more than a few GHz~\cite{Petta2005,Laird2006}. 

Here we demonstrate a radio-frequency reflectometry circuit, operating at $196~\text{MHz}$, that employs a SQUID as the primary amplifier~\cite{Muck2010a}.
We find an amplifier noise temperature below 600~mK.
This enables the reflectometry circuit to detect a capacitance signal with a sensitivity better than 0.1~$\text{aF}/\sqrt{\text{Hz}}$.
Attaching the reflectometry circuit to the ohmic contact of a GaAs double quantum dot, we acquire a Coulomb stability diagram with a resolution of $100 \times 100$ points within 20~ms.
The measurement time to distinguish the two states of a singlet-triplet qubit using gate-based capacitance sensing is estimated to be well below one microsecond even at low excitation power. This time should be short enough for single-shot readout.

This paper is organized as follows.
Section II describes the measurement setup and its principle of operation. 
In Section III we characterize the amplifier and show how to optimize its noise and gain by adjusting the control bias settings.
In Section IV, we tune the resonant tank circuit used in the reflectometry setup, and measure the capacitance sensitivity using a variable capacitor (varactor).
The setup used in Sections III-IV contains a single quantum dot device that serves as a realistic load on the tank circuit. %Transport measurements in the coulomb blockade  regime are presented in the supplementary information.
In Section V, this device is replaced by a double quantum dot, and the reflectometry circuit is used to measure the stability diagram and to estimate the minimum acquisition time. 
Section VI summarizes and evaluates the potential for qubit readout.
Further technical details and measurements of the single quantum dot appear in the Supplementary Information.

%Here, we demonstrate a radio-frequency reflectometry circuit operating at $196~\text{MHz}$ that employs a SQUID amplifier as the first stage of amplification~\cite{Muck2010a} combined with a tuneable impedance matching circuit. This paper is organized as follows: in part II we present the reflectometry circuit attached to a semiconductor single quantum dot. In part III and IV we use this setup to measure the SQUID amplifier's noise performance and tune the impedance of the circuit via a variable capacitor to find a capacitance sensitivity better than 0.1~$\text{aF}/\sqrt{\text{Hz}}$.  In part V, to further demonstrate the capabilities of our read-out circuit, we replace the previous device with a double quantum dot and find that we can acquire $100 \times 100$ points charge stability diagrams in only 20~ms. Finally, in part VI we discuss the results and estimate the dispersive read-out time of a singlet-triplet qubit through a gate electrode based on our capacitance sensitivity and input power. We find a read-out time of much less than a microsecond, which is faster than the relaxation time of the qubit and therefore enables single-shot read-out.  

\section{Measurement Setup}
\begin{figure}
\includegraphics[width=8.0cm]{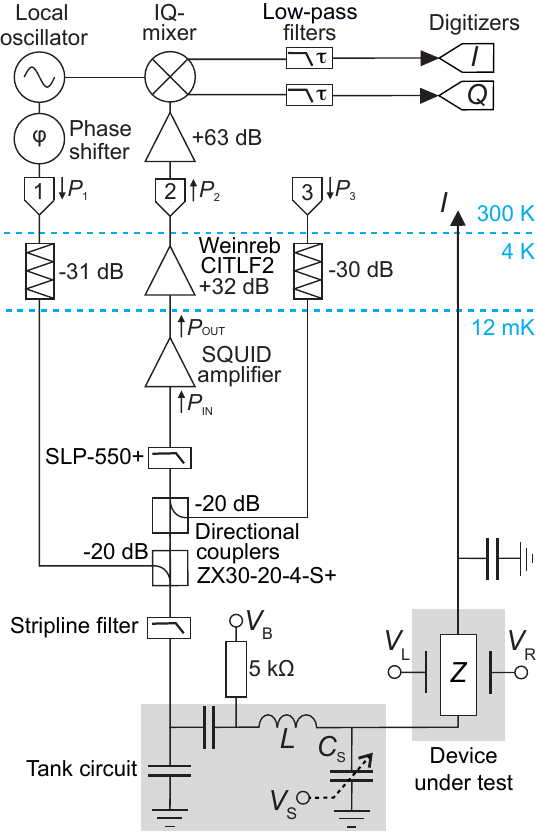}
\caption{Experimental setup. Measurements were performed in a dilution refrigerator at its base temperature of 12~mK.
A resonant tank circuit is defined by a surface mount inductor $L = 223$~nH and capacitors, including a varactor tuned by voltage $V_\text{S}$. 
 To excite this circuit, an RF carrier tone is generated by a local oscillator, phase shifted, injected into port~1 of the refrigerator with power $P_1$, and launched towards the tank circuit via cryogenic attenuators and a directional coupler. The reflected signal is amplified first by the SQUID and then by a semiconductor postamplifier, before it is fed via port~2 of the refrigerator to a homodyne mixing circuit to demodulate the signal into two voltages $\VI$ and $\VQ$ that represent the $I$ and $Q$ quadratures. Alternatively the output from port~2 is measured using a spectrum analyzer or network analyzer (not shown). A second injection path via port~3 is used to calibrate the amplifier chain.
 The tank circuit loaded by a device under test with impedance $Z$. For the experiments in Sections III-IV, $Z$ is a quantum dot fully pinched off using gate voltages $\VL$ and $\VR$.
 For the experiments in Section V, $Z$ is a double quantum dot.
\label{Fig:Schematic}}
\end{figure}

\begin{figure}
\includegraphics[width=8.5cm]{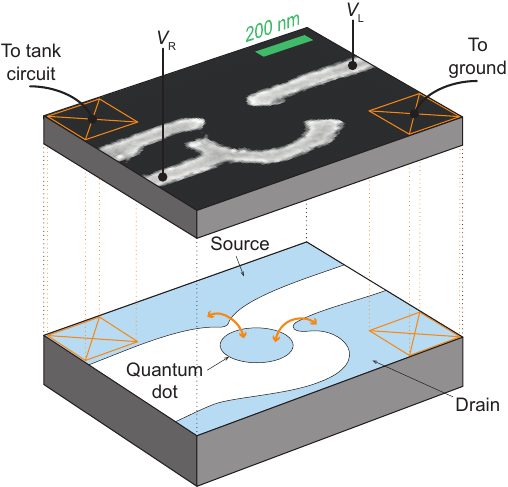}
\caption{Exploded plan of a GaAs quantum dot of the design used in this experiment~\cite{Mavalankar2013,Ares2016}.
The device consists of a GaAs/AlGaAs heterostructure wafer containing a two-dimensional electron gas 90\,nm beneath the surface.
On the surface of the wafer, Ti/Au gates are fabricated, as shown in the scanning electron microscope image (top).
Negative voltages applied to these gates selectively deplete the electron gas, defining source and drain regions and a quantum dot.
The source and drain are connected to the tank circuit and to ground by ohmic contacts, schematically indicated by squares.
Electron tunneling between source, dot, and drain contributes to the conductance and capacitance between the tank circuit and ground.
For the experiments in Sections III-IV, the dot is fully pinched off (i.e.\ electron tunneling is nearly suppressed) using very negative gate voltages~$\VL$ and~$\VR$.
The double quantum dot of Section V operates on the same principle.
\label{Fig:QuantumDot}}
\end{figure}

%To test the performance of the reflectometry circuit, we measure the sensitivity to two kinds of signal: changing capacitance in a varactor, and changing charge configuration in a GaAs quantum dot.
Figure~\ref{Fig:Schematic} shows the reflectometry circuit used in the experiment. It is designed to sensitively measure two kinds of signal: changing capacitance in a varactor, and changing charge configuration in a laterally defined quantum dot.
We have previously characterized these sensitivities using the same setup without the SQUID amplifier~\cite{Ares2016}.
Quantum dots of the kind used in this experiment are defined in a GaAs/AlGaAs two-dimensional electron gas (2DEG) by applying depletion voltages to top gates.
Here we use the device shown in Fig.~\ref{Fig:QuantumDot}, in which a quantum dot is defined by gate voltages $\VL$ and $\VR$, and measured via source and drain contacts to the 2DEG.
For the experiments of Sections III-IV this device was incorporated into the measurement circuit; however, in order to characterize the SQUID independently of the quantum dot, the dot was configured for very small conductance by applying large negative gate voltages and thereby also depleted of electrons.

 %The right gate was held at $\VR=-350\,\text{mV}$ while the left gate voltage $\VL$ was tuned to adjust the electrochemical potential on the dot.\\ 
The quantum dot device is wire-bonded to the RF circuit, that was assembled from chip components on a printed circuit board.
The RF circuit includes a fixed inductor $L$, a varactor of capacitance $C_\text{S}$ tuned by a voltage $V_\text{S}$, and a terminal through which a DC source-drain bias voltage $V_\text{B}$ is applied to the quantum dot device (Fig.~\ref{Fig:Schematic}).
These components form a resonant tank circuit with a total impedance that depends on the quantum dot impedance, the varactor tuning voltage, and the RF frequency.
The circuit board is mounted in a 12~mK dilution refrigerator wired for reflectometry measurements. An RF input line (port 1) injects power into the tank circuit via a directional coupler. The reflected signal is passed to a SQUID amplifier at base temperature, boosted by a semiconductor postamplifier at 4~K, and then measured at port 2. Once this amplifier chain is configured appropriately, its noise is dominated by the SQUID amplifier, which therefore sets the measurement sensitivity~\cite{Ares2016}. A second RF input line (port~3), coupled via an oppositely oriented directional coupler, allows calibrated signals to be injected directly into the RF measurement line to characterize the amplifier chain independently of the resonant circuit. Both input lines contain attenuators to suppress thermal noise.
\begin{figure}
\includegraphics[width=3.33in]{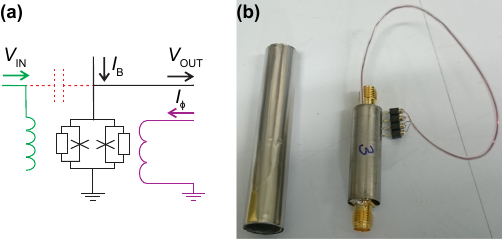}
\caption{(a) SQUID amplifier schematic, showing SQUID bias current $\IB$, flux bias current $\IFlux$, and input and output voltages $V_\text{IN}$ and $V_\text{OUT}$. The dashed capacitance between $V_\text{IN}$ and $V_\text{OUT}$ represents a parasitic capacitance when referring to the setup used in Sections III-IV. In the setup discussed in Section V, the dashed capacitor also represents an additional feedback capacitor, introduced to lower the input impedance of the amplifier. Each Josephson junction is shunted by a 30~$\Omega$ resistor.
(b) Photograph of a packaged amplifier (right) and its shield (left). Coaxial SMA connectors at bottom and top are RF input and output; the DC wires leading to a header connector are used to supply $I_\Phi$, $I_B$, a heating current, and DC ground.
Although the heater was not used in our experiment, it could be used to remove trapped flux by heating the SQUID above the critical temperature.\label{Fig:SQUIDschem}}
\end{figure}

The SQUID amplifier is shown schematically in Fig.~\ref{Fig:SQUIDschem}(a).
It exploits the fact that the critical current of a DC SQUID depends on the instantaneous magnetic flux enclosed between the junctions.
To operate the amplifier, an input signal $V_\text{IN}$ is fed into a 20-turn superconducting coil, acting as an open-ended transmission line, through which it excites an oscillating magnetic field. 
The coil is fabricated over a Nb-based SQUID, separated by a 400~nm $\text{SiO}_\text{2}$ spacer layer, following the `washer' geometry shown in Fig.~7(b) of Ref.~\cite{Muck2010a}.
The SQUID is biased with a current $\IB$ set greater than the maximum critical current. The resulting junction voltage $V_\text{OUT}$ depends on the instantaneous critical current, and its variation constitutes the amplifier's output signal~\cite{Muck2010a}.
To optimize the gain, % $dV_\text{OUT}/dV_\text{IN}$, 
a flux offset is applied by means of a flux bias current $\IFlux$ applied to a nearby coil.
The resonant frequency and quality factor of the input coil determine the optimum operating frequency and the bandwidth. The length of this input coil is chosen according to the desired operation frequency, since the gain peaks at a frequency that corresponds to approximately half a wavelength in the input coil~\cite{Muck2010a}.
The geometry was chosen to give an operating frequency near 200$\,$MHz with a bandwidth around 60$\,$MHz.
Figure~\ref{Fig:SQUIDschem}(b) shows a photograph of the amplifier with connectors.
The picture also shows a shield, made of lead and Conetic QQ foil, into which the amplifier is inserted to suppress flux noise.

For frequency-domain experiments, the output at port~2 is measured directly using a spectrum analyzer or a network analyzer.
For time-domain measurements, the signal is homodyne demodulated using a lock-in amplifier to yield in-phase and quadrature signals $I$ and $Q$, each filtered with time constant $\tau$ as in the circuit of Fig.~\ref{Fig:Schematic}.

%\section{SQUID amplifier performance}
\section{Characterizing and tuning the amplifier}

We begin by characterizing the amplifier's gain and noise. To achieve optimised signal-to-noise performance, we follow a tuning procedure that adjusts the current bias across the SQUID, the flux offset and input power.
For the measurements in this section, the amplifier is driven by direct injection into port 3, and the output from port 2 is measured using a spectrum analyzer (see Fig.~\ref{Fig:Schematic}). The injected tone has a frequency $f_\text{C}=196\,\text{MHz}$, chosen for later compatibility with the tank circuit.
For the measurements as a function of $\IB$ and $\IFlux$, we set the power at port 3 to be $P_3=-89\,\text{dBm}$, corresponding to a power $P_\text{IN}=-139\,\text{dBm}$ at the SQUID input. 
%This is well below the expected amplifier saturation power of $-100\,\text{dBm}$.
The SQUID amplifier gain is determined by comparing the total transmission from port~3 to port~2 with the amplifier present, versus an identical measurement in which it is replaced by a short length of cable. The gain is:
\begin{equation}
\frac{P_\text{OUT}}{P_\text{IN}} = \frac{|S_{32}|^2 (\text{amplifier present})}{|S_{32}|^2 (\text{amplifier absent})},
\end{equation}
where $P_\text{OUT}$ is the power at the amplifier output.
The noise power is then determined by injecting a signal tone with power $P_\text{IN}$ into the SQUID amplifier input, and measuring the output spectrum at port 2. The noise power referred to the amplifier input is then:
\begin{equation}
P_\text{N} = P_\text{IN} \frac{P_2 (\text{noise})}{P_2 (\text{signal})},
\end{equation}
where $P_2 (\text{signal})$ is the power of the amplified signal tone and $P_2 (\text{noise})$ is the noise power, both measured at port~2. The system noise power can then be expressed as a noise temperature
\begin{equation}
T_\text{N} = \frac{P_\text{N}}{k_\text{B} \Delta f} - T,
\label{eq:Tn}
\end{equation}
where $\Delta f$ is the resolution bandwidth of the spectrum analyzer and $T$ is the noise temperature of the input signal into the SQUID amplifier.
To accurately determine the power level $P_\text{IN}$, which depends on the transmission characteristics of the cables, we separately measured the attenuation of the injection path (see Supplementary Information).
%Possible reflections on the amplifier input due to impedance mismatches and the exact temperature $T$ associated with the input signal are not known.
To avoid underestimating the system noise temperature, we assume the lowest possible electron temperature given perfect thermalization, i.e.\ $T=12\,\text{mK}$. 
(In fact, our typical estimated electron temperature is $T \approx 25$\,mK~\cite{Mavalankar2016}.)
This assumption and the possibility of reflections on the amplifier input make $T_\text{N}$ an upper limit to the noise temperature of the SQUID amplifier.

Optimum operation, i.e.\ high amplifier gain and low noise, requires us to find suitable settings for both $\IB$ and $\IFlux$. We follow a two-step process. First we increase $\IB$ until a change in $P_\text{OUT}$ is detected, indicating that the critical current has been exceeded. Next we optimize the flux offset $\Phi$ via $\IFlux$ to find a steep point in the function $V_\text{OUT}(\Phi)$, so that the output voltage is most sensitive to the induced flux.
%in $V_\text{OUT}(\Phi+\delta\Phi)$, where the voltage drop is most sensitive to the flux oscillations from the RF input signal $\delta\Phi$. 
\begin{figure}[t!]
\includegraphics[width=3.33in]{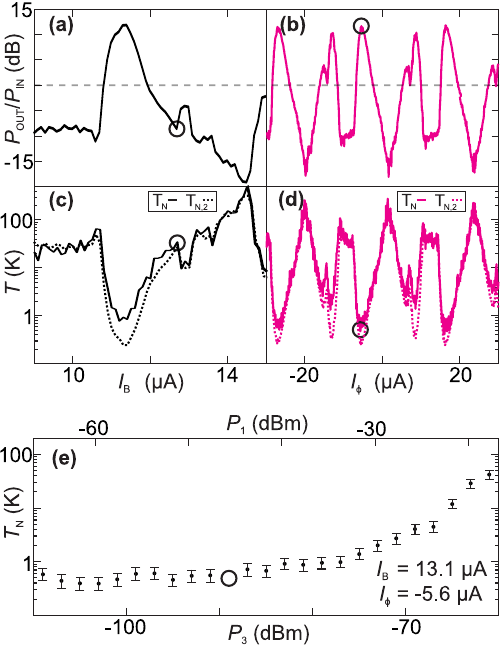}
\caption{(a),(b),(c) and (d) show the characterization of the SQUID amplifier as a function of the current bias $\IB$ (black) and the flux coil bias $\IFlux$ (purple) at frequency $196\,\text{MHz}$ and power into port 3 $P_3 = -89\,\text{dBm}$. (a) Gain as a function of bias, with $\IFlux=0$. The grey dashed line marks $0\,\text{dB}$. (b) Gain as a function of flux bias current at $\IB=13.1\,\upmu\text{A}$. (c) Noise temperature $T_\text{N}$ as a function of $\IB$ determined for every point in (a). The dashed line is the postamplifier contribution $T_\text{N,2}$. (d) Noise temperatures $T_\text{N}$ (line) and $T_\text{N,2}$ (dashed line) as a function of $\IFlux$. The black markers indicate the chosen settings ($\IB=13.1\,\upmu\text{A}$ and $\IFlux=-5.6\,\upmu\text{A}$) in the rest of Sections III-IV, giving gain $\approx 12\,\text{dB}$ and $T_\text{N}\,\approx480\,\text{mK}$. (e) Noise temperature as a function of input power into port 3 $P_3$ (bottom axis).
%measured at $I_{\Phi}=-5.6\,\upmu\text{A}$ and $I_\text{B}=13.1\,\upmu\text{A}$.
The top axis shows the estimated corresponding input power into port 1 $P_1$, assuming $|S_{21}|=-49.98\,\text{dB}$ from the best matching condition in Fig.~\ref{Fig:Sc}. The black circle indicates the power used in (a), (b), (c) and (d). \label{Fig:SQUID}}
\end{figure}

Figure~\ref{Fig:SQUID} shows the performance of the amplifier as a function of the bias currents $\IB$ (Fig.~\ref{Fig:SQUID}(a),(c)) and $\IFlux$ (Fig.~\ref{Fig:SQUID}(b),(d)). At low $\IB$, the SQUID is biased below its critical current and only a fraction of the input power is transmitted to the output by capacitive leakage. As $\IB$ is increased above the critical current a voltage develops and the gain increases abruptly.
 This occurs at $\IB \approx 10.7\,\upmu\text{A}$ in Fig.~\ref{Fig:SQUID}(a).
At larger currents, the gain varies non-monotonically  due to the self-inductance of the SQUID~\cite{Clarke2006}. These variations can be compensated by adjusting $\IFlux$, and in fact we find that a similar gain can be achieved for all chosen values of $\IB$ larger than the critical current. Because the critical current depends on the flux, the chosen $\IB$ should be larger than the critical current for all $\IFlux$.

We now measure the gain as a function of flux bias current $\IFlux$ (Fig.~\ref{Fig:SQUID}(b)).
For this measurement, we choose $\IB=13.1\,\upmu\text{A}$ (black marker in Fig.~\ref{Fig:SQUID}(a) and (c)).
At first sight, Fig.~\ref{Fig:SQUID}(a) implies that $\IB$ is larger than optimal; the reason to choose this value is that on a previous cooldown, the critical current was as high as $12.9\,\upmu\text{A}$ at $I_\Phi = 0$ (see Supplementary Information).
By choosing $\IB$ above this value, we aim for it to be well above the critical current for all flux-offsets but not large enough to significantly heat the SQUID.

As shown in Fig.~\ref{Fig:SQUID}(b), the gain varies periodically with $\IFlux$, reflecting the periodic dependence of critical current on flux.
For an ideal SQUID at high current bias, the gain would be a sinusoidal function of flux. 
In fact, this amplifier has a more complex periodic dependence, which indicates that self-heating, junction asymmetry, and/or parasitic impedances play important roles in determining the gain~\cite{Clarke2006}.
For example, junction asymmetry would unequally divide the bias current between the two arms of the SQUID, leading to a changing flux.
To optimize the sensitivity we choose $\IFlux=-5.6\,\upmu\text{A}$, (black marker in Fig.~\ref{Fig:SQUID}(b) and (d)), leading to a gain of $11.7\,\pm0.8$ dB. 
The uncertainty of this value is accumulated over multiple measurements that are needed to determine the losses of the insertion path and the gain of the postamplifier.

Figures~\ref{Fig:SQUID}(c) and~\ref{Fig:SQUID}(d) show the system noise temperature $T_\text{N}$ as a function of $\IB$ and of $\IFlux$ respectively. In both traces, the same bias settings that maximize the gain also lead to low noise. To distinguish the noise of the SQUID from the noise of the postamplifier, we plot as a dashed curve on the same axes the postamplifier's contribution to the system noise temperature
\begin{equation}
 T_\text{N,2} = T_\text{P} \frac{P_\text{IN}}{P_\text{OUT}},  
\label{eq:Tn2}
\end{equation}
where $T_\text{P}= 3.7$~K is the input noise temperature of the postamplifier. This is the lowest noise temperature (referred to the SQUID input) that the system could achieve if the SQUID were a noiseless amplifier. Over most of the range, this contribution is approximately equal to the entire system noise ($T_\text{N} \approx T_\text{N,2}$), meaning that the intrinsic noise of the SQUID is indeed undetectable. However, the optimal bias settings, with highest gain, lowest noise and therefore best signal-to-noise ratio, lead to $T_\text{N}>T_\text{N,2}$, showing that for these settings the system noise is dominated by the SQUID contribution. Previous experiments have found this contribution to arise from hot electrons generated by ohmic dissipation~\cite{Muck2010a,Wellstood1994,Muck2001}. There may also be a contribution from thermal radiation leaking into the SQUID. The lowest noise temperature observed is $T_\text{N}=500\pm100\,\text{mK}$,
obtained with $\IFlux=-5.6\,\upmu\text{A}$ (black marker in Fig.~\ref{Fig:SQUID}(d)). This is within a factor 120 
of the quantum limit $hf_\text{C}/2k_\text{B} = 5$~mK~\cite{Clerk2010}.

\begin{figure}[t!]
\includegraphics[width=3.33in]{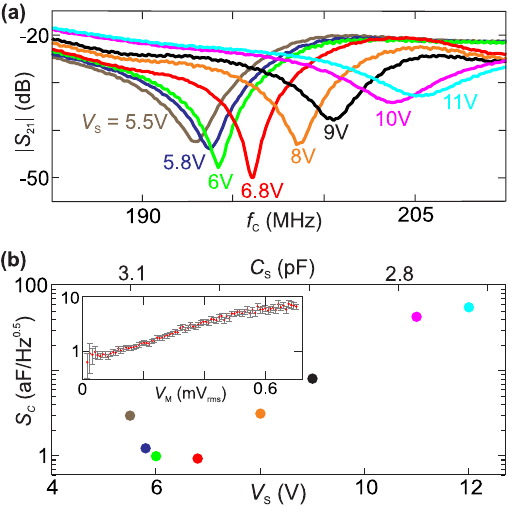}
\caption{(a) Transmission $|S_{21}|$ from port 1 to port 2 as a function of carrier frequency $f_\text{C}$ at the varactor voltage settings indicated. (b) Capacitance sensitivity $S_C$ as a function of varactor voltage $V_\text{S}$, measured with a modulation frequency $f_\text{M}=3\,\text{kHz}$, modulation amplitude $V_\text{M}=99\,\upmu\text{V}_\text{rms}$ (corresponding to a capacitance modulation of $\delta C=6.7\,\text{aF}_\text{rms}$) and carrier power $P_1=-60\,\text{dBm}$. The carrier frequency was adjusted to the best matching point for each setting of $V_\text{S}$. Inset: Capacitance sensitivity $S_C$ as a function of $V_\text{M}$ at optimal matching ($V_\text{S}=6.8\,\text{V}$, $f_\text{C}=196\,\text{MHz}$). The error bars derive from the height of the signal sideband compared to the scatter in the noise background.\label{Fig:Sc}}
\end{figure}

To study the amplifier dynamic range, Fig.(e) shows the noise temperature %at $I_{\Phi}=-5.6\,\upmu\text{A}$
as a function of input power $P_3$. The top axis shows an estimate of the corresponding power $P_1$ into port 1 that leads to the same power at the SQUID input when it is used for a reflectometry experiment (assuming that the matching circuit is optimized, as discussed below in Sec.~\ref{Sec:matching} and Fig.~\ref{Fig:Sc}).
The noise increases at high input power, with the threshold being approximately $P_3 \approx -70~\text{dBm}$, which corresponds to an amplifier input power of approximately
$P_\text{IN}\approx -120\,\text{dBm}$.
The input power corresponding to the onset of amplifier saturation can be roughly estimated from the SQUID parameters given by the manufacturer to be $P_\text{IN}\approx-100\,\text{dBm}$ (see Supplementary).
The lower dynamic range of the amplifier in our setup  and the elevated noise temperature (compared to the state of the art in setups dedicated to optimised SQUID amplifier performance rather than sensitive RF read-out) could be related to poor input impedance matching between the SQUID and the $50\,\Omega$ components in the circuit, to radiation from outside the refrigerator, or to incomplete thermalization~\cite{Muck2010a}.

\section{Optimizing the capacitance sensitivity\label{Sec:Scopti}}
We now show how to use the amplifier for sensitive measurements of capacitance. These measurements use a reflectometry configuration, in which the signal is injected via port~1 and the reflected signal is amplified by the SQUID. To avoid any contribution from the quantum capacitance, gate voltages are set to completely empty the quantum dot. To perform these measurements, we first tune the impedance of the tank circuit close to that of the measurement circuit, and then characterize the sensitivity to changes in the capacitance~\cite{Ares2016}.

The capacitance sensitivity $S_C$ is determined by modulating the varactor capacitance at a frequency $f_\text{M}$ while driving the tank circuit at carrier frequency $f_\text{C}$.
The reflected signal, monitored at port 2 using a spectrum analyser, contains a main peak at $f_\text{C}$ and sidebands at $f_\text{C}\pm f_\text{M}$. Such sidebands arise from mixing of an amplitude-modulated output signal when the impedance of the resonant circuit is sensitive to the modulated quantity. $S_C$ is extracted from the height of the sidebands above the noise floor (i.e.\ the signal-to-noise ratio or SNR, expressed in dB) according to~\cite{Ares2016,Aassime2001,Brenning2006}:\\
\begin{equation}
S_C=\frac{\delta C}{\sqrt{2\Delta f}}10^{-\text{SNR}/20}\label{Eq:SB}
\end{equation}
where $\Delta f$ is the spectrum resolution bandwidth and $\delta C$ the root-mean-square modulation amplitude of the capacitance.
To generate a capacitance modulation, we vary the control voltage of the varactor $V_\text{S}$ with amplitude $V_\text{M}$, which is converted to the capacitance modulation $\delta C$ as explained in the Supplementary Information.

\subsection{Optimizing the matching circuit\label{Sec:matching}}
To optimize the impedance matching between the tank circuit and the input network, we tune the varactor using $V_\text{S}$. Figure~\ref{Fig:Sc}(a) shows the transmission $|S_{21}|$ from port~1 to port~2, which is proportional to the tank circuit's reflection coefficient, for different settings of $V_\text{S}$. The lowest reflection coefficient, and therefore the best match, is achieved at $f_\text{C} = 196\,\text{MHz}$ when $V_\text{S} = 6.8\,\text{V}$.

Figure~\ref{Fig:Sc}(b) shows the capacitance sensitivity as a function of $V_\text{S}$ measured with an input power of $P_1=-60\,\text{dBm}$ into port~1. This power corresponds to approximately $-154\,\text{dBm}$ on the SQUID input and is well below the threshold of amplifier saturation. The best sensitivity is $S_C=0.9\pm0.2\,\text{aF}/\sqrt{\text{Hz}}$. As expected, this occurs closest to perfect matching and therefore this varactor setting with the associated resonance frequency of $196\,\text{MHz}$ is used in the remainder of Sec.~\ref{Sec:Scopti}~\cite{Ares2016}.

The inset of Fig.~\ref{Fig:Sc}(b) is a plot of the sensitivity as a function of modulation amplitude $V_\text{M}$, measured using the optimized matching parameters. These data show that the sensitivity degrades at high modulation amplitude due to non-linearity of the varactor, but confirm that the modulation applied in the main panel, $V_\text{M}=99\,\upmu\text{V}_\text{rms}$, is within the linear range. In the following measurements (Sec.~\ref{Sec:ScoptiB}) we choose an even smaller modulation amplitude of $V_\text{M} = 80\,\upmu\text{V}_\text{rms}$.

\subsection{Optimizing the input power}
\label{Sec:ScoptiB}
\begin{figure}[tb]
\includegraphics[width=3.33in]{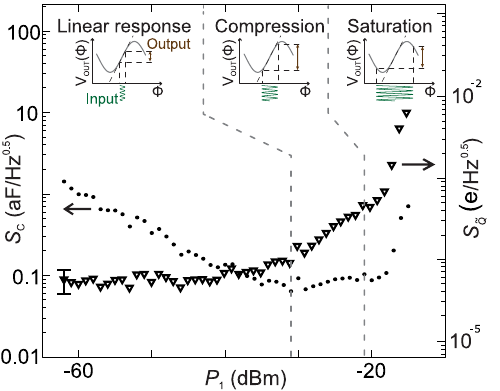}
\caption{Capacitance sensitivity $S_C$ (left axis, points) and the sensitivity to a charge on one plate of the varactor $S_{\tilde{Q}}$ (right axis, triangles) as a function of the carrier power at port 1 $P_1$. The errors in $S_C$ are smaller than the symbols and due to uncertainties in determining the noise level. The errors in $S_{\tilde{Q}}$ are due to uncertainties from the noise level as well as the input lines/cables. For clarity only one error bar is marked. Other parameters: $f_\text{M}=3\,\text{kHz}$, $V_\text{M}= 80\,\upmu\text{V}_\text{rms}$, $f_\text{C}=196\,\text{MHz}$ and $V_\text{S}=6.8\,\text{V}$. The insets illustrate the three operating regimes (see text) by marking the input and output signals on a graph of the flux-to-voltage transfer function $V_\text{OUT}(\Phi)$.\label{Fig:SCV0}}
\end{figure}

Next we study how the capacitance sensitivity depends on the carrier power $P_1$. Figure~\ref{Fig:SCV0} shows that increasing $P_1$ improves the sensitivity, up to an optimal power of $P_1=-31\,\text{dBm}$, where the sensitivity reaches $S_C=0.07\pm0.02\,\text{aF}/\sqrt{\text{Hz}}$. This power corresponds to
approximately $-125\,\text{dBm}$ incident on the amplifier input, given the known losses due to attenuation and reflection on the tank circuit following the signal path associated with input port 1.
From $-31\,\text{dBm}$ to around $-21\,\text{dBm}$ the sensitivity stays roughly constant before worsening at higher input powers.

We interpret these three regimes using the flux-to-voltage transfer function of the SQUID $V_\text{OUT}(\Phi)$, as indicated by the insets in Fig.~\ref{Fig:SCV0}. For $P_1 < -31$~dBm, the amplifier is in its linear-response regime where the gain and the noise temperature are constant such that the sensitivity improves with increasing SNR at increasing input power. The region of approximately constant sensitivity between $-31$~dBm and $-21$~dBm indicates gain compression, which means that the flux $\delta\Phi$ induced by the input signal exceeds the linear range of $V_\text{OUT}(\Phi+\delta\Phi)$. 
This creates harmonics sidebands in the output spectrum such that the SNR around the main sidebands decreases.
For $P_1>-21\,\text{dBm}$, when $\delta \Phi$ exceeds a quarter of a flux period, the amplifier reaches its saturation. At this point the flux oscillation reaches beyond the maxima and minima of $V_\text{OUT}(\Phi)$ and the sensitivity is degraded. The saturation threshold in Fig.~\ref{Fig:SCV0} approximately matches the power threshold where $T_\text{N}$  begins to worsen (Fig.~\ref{Fig:SQUID}(e)). $S_C$ does not follow the noise temperature exactly because increasing the carrier power affects both the signal and the noise.
In the next paragraph we will introduce a figure of merit that does not benefit from input power and follows the noise more closely.

For dispersive readout of spin qubits, good capacitance sensitivity $S_C$ is not sufficient to achieve high fidelity. One reason is that it may require a large RF bias, giving rise to back action by exciting unwanted transitions in the qubit device. Another reason is that the quantum capacitance is usually sizable only within a small bias range, so that increasing the RF excitation improves $S_C$ without improving the qubit readout fidelity. This is the case for singlet-triplet qubits, where the quantum capacitance is large only near zero detuning~\cite{Petersson2010}.
As explained in the Supplementary Information, for dispersive readout the crucial sensitivity is to the oscillating charge induced on the gate electrode by the qubit capacitance, which in our setup corresponds to the charge induced on one plate of the varactor. This sensitivity is
\begin{equation}
S_{\Qtilde}=\sqrt{2}V_0 S_C
\end{equation}
where $V_0$ is the root-mean-square RF voltage across the device~\cite{Ares2016}.
This is a key figure of merit for dispersive spin qubit readout.
For single-shot readout, this sensitivity must allow for detecting a charge smaller than one electron within the qubit lifetime. We estimate $V_0$ using a circuit model of the tank circuit as in Ref.~\cite{Ares2016}. For example, at $P_1 = -29\,\text{dBm}$, the incident power onto the tank circuit is $\sim10\,\text{pW}$, giving an estimated voltage $V_0 = 192\,\upmu\text{V}_{\text{rms}}$ across the device.
The right axis of Fig.~\ref{Fig:SCV0} shows
$S_{\Qtilde}$ as a function of input power into port~1.
$S_{\Qtilde}$ worsens at slightly lower input power than does $S_C$, but reaches below $ 100\,\upmu e/\sqrt{\text{Hz}}$ for optimal settings.

\section{Fast readout of a double quantum dot}

To demonstrate the full functionality of the circuit, we measure charge stability diagrams and determine the acquisition rate. We replace the single quantum dot from Fig.~\ref{Fig:Schematic} with a double quantum dot operated in the Coulomb blockade regime~\cite{camenzind2019}.
For this experiment we use a different SQUID amplifier which includes a feedback capacitor between  $V_\text{IN}$ and  $V_\text{OUT}$, designed to lower its input impedance and thus improve the matching to the $50$\,$\Omega$ line impedance.
We drive our circuit at $f_\text{C}=210\,\text{MHz}$ with a power of  $P_1=-35\,\text{dBm}$, which is just below the threshold for broadening the Coulomb peaks in the double dot. To form the double dot, we automatically adjust the gate voltages with the help of machine-learning algorithm~\cite{Moon2020}.

\begin{figure*}[htb]
\includegraphics[width=7in]{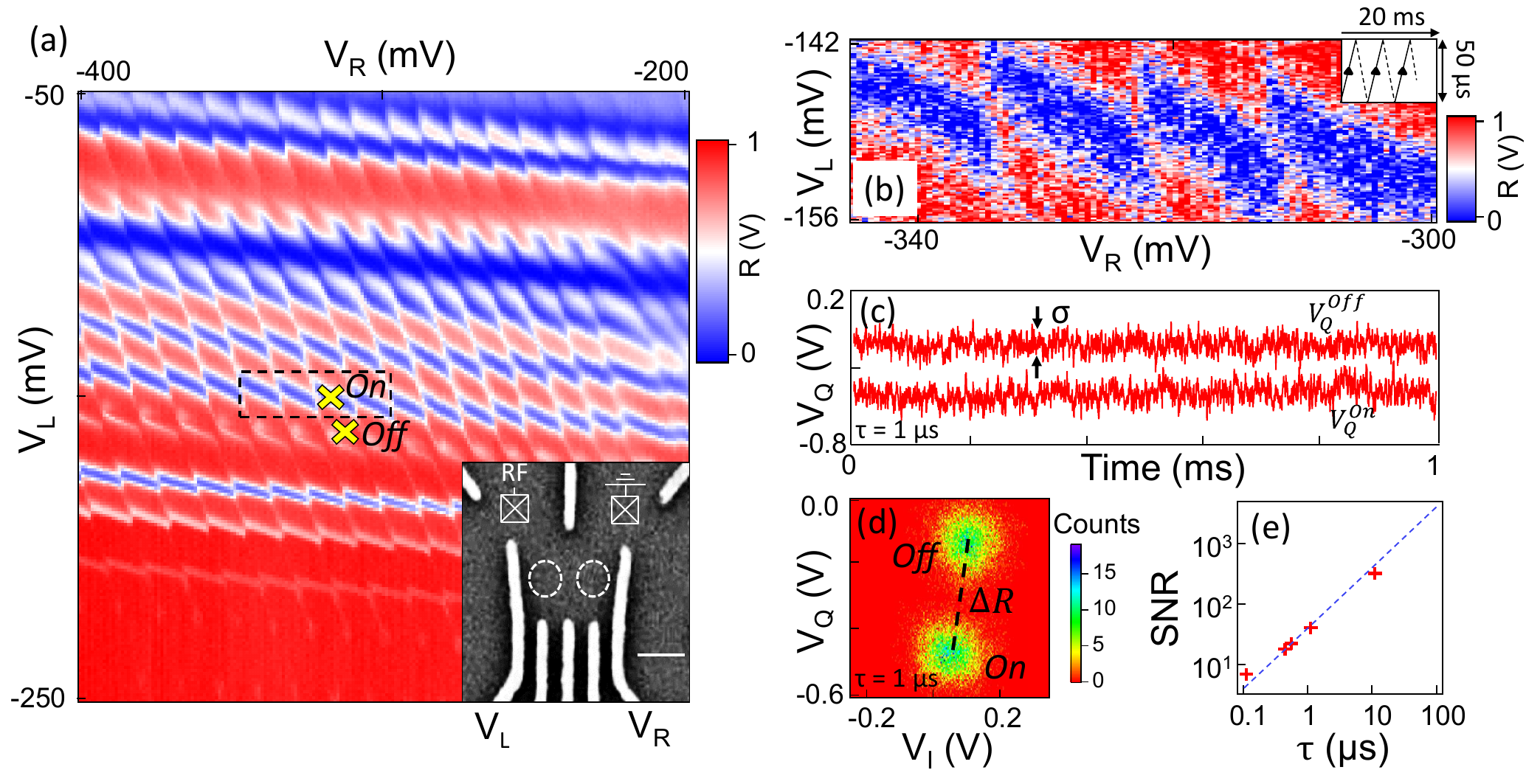}
\caption{(a) Reflected amplitude showing the double quantum dot charge stability diagram. The amplitude is normalized from 0 to 1 based on the minimum and maximum measured value.  Inset: SEM image of the device. The scale bar is 200\,nm long and the two dashed circles symbolise the two quantum dots.  (b) Fast measurement of the charge stability diagram area highlighted by the dashed rectangular in (a) obtained with $\tau = 1\,\mu\text{s}$. $\VL$ and $\VR$ are swept using triangular waveforms as illustrated in the inset. (c) Time trace of normalized $\VQ$ measured at the \emph{On} and \emph{Off} coordinates of (a), $\VQ^\text{On}$ and $\VQ^\text{Off}$ respectively. $\sigma$ is the standard deviation of the trace. (d) Joint histogram of recorded $\VI$ and $\VQ$ values for \emph{On} and \emph{Off}. (e) SNR as function of $\tau$ (symbols) and fit to Eq.~(\ref{Equ:Tau}) with $\tau_\text{min} = 25\,\text{ns}$ (line).}\label{Fig:DQD}
\end{figure*}

The charge stability diagram of the double dot is shown in Fig.~\ref{Fig:DQD}(a), which plots the normalised signal amplitude $R=\left | \VI+i\VQ \right |$ as a function of the left and right plunger voltages $\VL$ and $\VR$. 
This plot shows the characteristic honeycomb pattern of a double quantum dot.
In the centre of each honeycomb, Coulomb blockade suppresses conductance, and the reflected signal is large (red regions in  Fig.~\ref{Fig:DQD} (a)); at the honeycomb boundaries, Coulomb blockade is partly lifted and the signal is small (blue regions in  Fig.~\ref{Fig:DQD} (a)) \cite {VanderWiel2002, Stehlik2015}.
As expected, the charge transitions of the left dot, closer to the RF electrode, give the strongest signal.

%When the conductance of the double quantum dot is suppressed by Coulomb blockade, the amplitude of the reflected signal is at a maximum (red regions in  Fig.~\ref{Fig:DQD} (a)).
%Conversely when the gate voltages are set  at a charge transition of the left dot, for which electron tunneling can occur, the reflected amplitude is lower (blue regions in  Fig.~\ref{Fig:DQD} (a)) \cite {VanderWiel2002, Stehlik2015}.

The low noise of the SQUID amplifier allows rapid measurement of the stability diagram.
To show this, we focus on the region of the stability diagram  marked by a dashed box in Fig.~\ref{Fig:DQD}(a), and apply triangular waveforms via on-board bias tees to rapidly sweep $\VL$ and $\VR$ over this range.
For these data, the filter time constant is set to $\tau = 1\,\mu\text{s}$.
We record $R$ during the upward ramps of the fast triangular waveform in order to build up a two-dimensional map (Fig.~\ref{Fig:DQD}(b)).
The resolution is $100 \times 100$ data points and the digitizer sample rate is $1\,\text{MHz}$, meaning that the entire plot is acquired within 20\,ms.
As expected, the resulting charge stability diagram, presented in Fig.~\ref{Fig:DQD}(b), shows the same pattern as in Fig.~\ref{Fig:DQD}(a), with easily distinguishable charge transitions despite the very short acquisition time.

The SNR can now be extracted directly by comparing a signal amplitude to the noise recorded in a time trace.
The signal in this case is taken as the difference in reflected power between gate configurations on and off a Coulomb peak.
To measure the SNR in Fig.~\ref{Fig:DQD}(e) we record $10,000$ samples of $\VI$ and $\VQ$, digitized at a rate of $10\,\text{MHz}$, at two locations in the charge stability diagram in Fig.~\ref{Fig:DQD}(a); $\VI^\text{On}$ and $\VQ^\text{On}$ at location marked by \emph{On} and $\VI^\text{Off}$ and $\VQ^\text{Off}$ at location marked by \emph{Off}.
This experiment is repeated for different choices of filter time constant $\tau$.
A typical pair of time traces is shown in Fig.~\ref{Fig:DQD}(c).
Figure~\ref{Fig:DQD}(d) represents these data as a joint histogram in the $\VI$ $\VQ$ quadrature space.
The two well-separated Gaussian distributions show that the two Coulomb states can be distinguished within a single time interval of duration~$\tau = 1~\mu$s.
The amplitude of the signal $\Delta R$ is defined as the distance between the mean values of the two distributions $\Delta R^2= \left [ \left \langle \VI^\text{on} \right \rangle - \left \langle \VI^\text{off} \right \rangle \right ]^2 + \left [ \left \langle \VQ^\text{on} \right \rangle - \left \langle \VQ^\text{off} \right \rangle \right ]^2$ and the noise $\sigma$ is their standard deviation (which as expected is the same in both $\VI$ and $\VQ$ channels).
The signal-to-noise ratio $\text{SNR} =\Delta R^2/\sigma^2$ is plotted as function of $\tau$ in Fig.~\ref{Fig:DQD}(e). The blue dashed line is a fit according to
\begin{equation}
\text{SNR}= \frac{\tau}{\tau_{\text{min}}}
\label{Equ:Tau}
\end{equation}
with fit parameter $\tau_{\text{min}} = 25\,\text{ns}$, which is the extrapolated time to distinguish the two configurations with SNR of unity.
The point at $\tau=100\,\text{ns}$ falls slightly above the fit line because the integration time of the digital converter ($100\,\text{ns}$) adds extra averaging. From these data, the sensitivity to a quasi-static charge of the double-dot device is at least as good as
\begin{equation}
%    S_{\DeltaQ} \leq e\sqrt{\tau_\text{min}} = 158\,\upmu e/\sqrt{\text{Hz}}.
    S_{\DeltaQ} \leq e\sqrt{\tau_\text{min}} = 160\,\upmu e/\sqrt{\text{Hz}}.
    \label{eq:SDeltaQ}
\end{equation}
In this expression, $\DeltaQ$ is the difference in charge induced on the quantum dot between the two measured configurations, which is a large fraction of one electron charge.

In the Supplementary Information we present a measurement of the sensitivity to a small charge modulation $\delta Q$, measured on the steep flank of a Coulomb peak using a single-dot device. This leads to a somewhat better sensitivity, but is not directly comparable because it was measured using a different amplifier. Both these charge sensitivities are distinct from the sensitivity $S_{\Qtilde}$ plotted in Fig.~4; $\Qtilde$ is a charge oscillating in response to the RF field, whereas $\DeltaQ$ and $\deltaQ$ are quasistatic charges. The former is what is measured in a dispersive measurement, the latter are what is measured using a charge sensor.

\section{Discussion}
We have shown that radio-frequency measurements using a SQUID amplifier can attain much better sensitivity than using a cryogenic semiconductor amplifier alone. This advantage holds when the signal level is limited by the need to avoid back-action on the device being measured, which is nearly always the case for quantum devices. The SQUID measured here has a gain around 12~dB and reaches a noise temperature below 600~mK, which is approximately 7 times better than the (already optimized) semiconductor amplifier. When used to measure capacitance via radio-frequency reflectometry, it allows a record capacitance sensitivity of $S_C=0.07\pm0.02\,\text{aF}/\sqrt{\text{Hz}}$, which corresponds to an improvement by a factor of 23 compared with the same setup without the SQUID~\cite{Ares2016}. This setup can also be used with a single quantum dot charge sensor. In the Supplementary Information, we perform this measurement and find a charge sensitivity  of  $S_{\deltaQ}=60\pm20\,\upmu e/\sqrt{\text{Hz}}$, corresponding to an improvement by a factor of 27 compared to the setup without the SQUID~\cite{Ares2016}. This improvement is better than expected from the improved noise temperature alone, and probably also arises from lower cable loss and a different impedance matching condition to the amplifier input.

To put these results in the context of spin qubit readout, we estimate the dispersive read-out time in a singlet-triplet qubit with the RF circuit connected to a plunger gate. In this case, a difference in capacitance on the order of 2~fF~ needs to be resolved to determine the state of the qubit \cite{Petersson2010}. Based on the capacitance sensitivity obtained in section \ref{Sec:Scopti} we estimate a single-shot readout time of $\sim 26~\text{ns}$ with our circuit (see Supplementary Information). Integrating a SQUID amplifier into a spin qubit setup should therefore significantly reduce the measurement noise, ultimately improving single-shot readout fidelity. This represents a major advantage for scalable quantum information processing architectures containing many qubits in a small space~\cite{Veldhorst2016a,Vandersypen2017}.

%allowing the single-shot readout threshold to be surpassed while operating in the radio-frequency range. This represents a major advantage for scalable quantum information processing architectures containing many qubits in a small space~\cite{Veldhorst2016a,Vandersypen2017}.

As well as for qubit readout, this setup can also increase the sensitivity of other radio-frequency measurements. The fast measurements of a double quantum dot presented above demonstrate a minimum per-pixel integration time $\tau_\text{min} \approx 25~\text{ns}$. This integration time is of the same order as the integration times in double-quantum dot measurements using Josephson parametric amplifiers~\cite{Stehlik2015, Morton2019} or high quality-factor microwave resonators~\cite{Vandersypen2019}, but was  enabled by a commercially available amplifier without the need for a dedicated fabrication environment. In another application of our circuit, the improved sensitivity provided by the SQUID has enabled time-resolved measurements of a vibrating carbon nanotube transistor~\cite{Wen2019}. 

\section*{Supplementary Material}
See supplementary material for more explanation of the charge sensitivity, charge sensing measurements on a single quantum dot, data from a separate cool-down of the amplifier, details of the measurement calibration, and full instructions for installing and tuning the amplifier.

\section*{Acknowledgements}
The work was funded by DSTL (contract 1415Nat-PhD\_59),
EPSRC (EP/J015067/1, EP/N014995/1),
the Royal Academy of Engineering,
a Marie Curie Fellowship,
Templeton World Charity Foundation,
and the the European Research Council (grant agreement 818751).
We thank M. M\"{u}ck and ez SQUID for supplying and assisting with the amplifiers.

The data that support the findings of this study are available from the corresponding author
upon reasonable request.
\vfill

\end{document}